# Content Based Image Retrieval System Using NOHIS-tree

Mounira TAILEB
King Abdulaziz University
Faculty of Computing and Information Technology
P.O Box 42808
Jeddah. 21551
mtaileb@kau.edu.sa

## ABSTRACT
Content-based image retrieval (CBIR) has been one of the most important research areas in computer vision. It is a widely used method for searching images in huge databases. In this paper we present a CBIR system called NOHIS-Search. The system is based on the indexing technique NOHIS-tree. The two phases of the system are described and the performance of the system is illustrated with the image database *ImagEval*. NOHIS-Search system was compared to other two CBIR systems; the first that using PDDP indexing algorithm and the second system is that using the sequential search. Results show that NOHIS-Search system outperforms the two other systems.

## Categories and Subject Descriptors
H.3.3 [**Information Storage and Retrieval**]: Information Search and Retrieval – *clustering, search process.*

## 1. INTRODUCTION
The increasing size of image databases requires the use of image retrieval systems. There are two approaches in such systems, text-based and content-based image retrieval (CBIR). In text-based approach images are indexed by text terms and retrieved by matching terms in query with those indexed. However, text annotation is inadequate with large databases. The content-based retrieval systems have become increasingly important in many applications areas such as geography, commerce, medicine. CBIR systems retrieve the most similar images to a query within an image database. This process requires describing automatically the visual content of the images and represents each image by a set of multidimensional vectors called descriptors [1].

In this paper a content-based image retrieval system called NOHIS-Search is described with its two phases. The rest of the paper is organized as follows. Section 2 describes the two phase of NOHIS-Search. In the off-line phase the indexing method is detailed and in the on-line phase the search algorithm is given. Section 3 evaluates the performance of NOHIS-Search and finally, the section 4 concludes the paper.

## 2. ARCHITECTURE OF NOHIS-SEARCH
The architecture of the implemented CBIR system, NOHIS-Search, is presented in figure 1. It consists mainly of two phases; the off-line and on-line phase. In the off-line phase the descriptors are extracted from each image in the database, and then the descriptors are indexed using the high-dimensional indexing method NOHIS-tree [2]. The on-line phase handles the process of querying the image database; the descriptors of the query are extracted and compared with that of other images in the database. The images that are visually similar to the query are displayed.

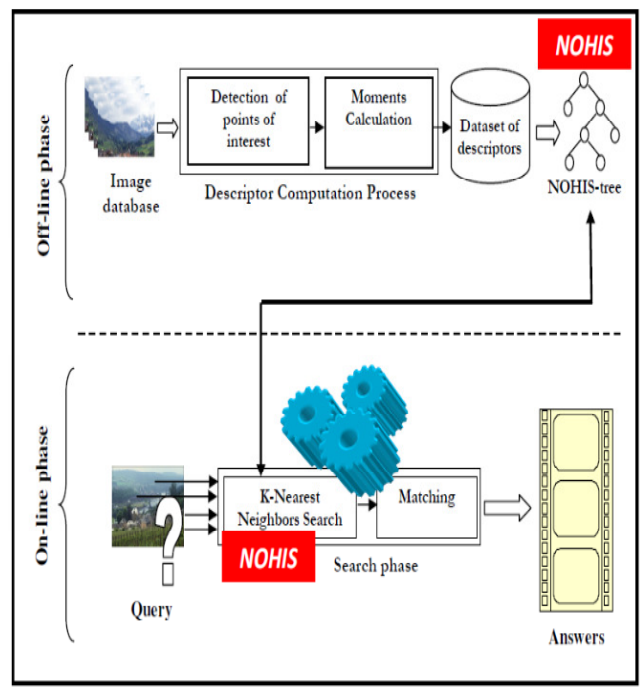

Figure 1. Architecture of NOHIS-Search

## 2.1 Off-line phase
### 2.1.1 Descriptor computation
In the content-based image retrieval systems descriptor computation is an important step. There are globally two types of descriptors; global and local descriptor. In NOHIS-Search, the image indexing technique used is based on the computation of local descriptors around the points of interest. We chose the multi-scale Harris detector [3] based on the Harris detector [4] following a comparative study between several points of interest in [5].

To compute the descriptors, regions around the points of interest are characterized using Zernike moments. These descriptors are invariant to rotation and fast in computation time. In [5], Zernike moments were compared with Hu moments and the moments of Legendre, Zernike moments are more efficient in indexing than the two other moments. They are also invariant to rotation, translation, change of scale and they are robust in the case of noisy images.

The used descriptors are the coefficients of the Zernike moments of order 3. The dimension of the descriptor is 12.

### 2.1.2 The Indexing technique NOHIS-tree

The existing multi-dimensional indexing techniques can be divided in two groups according to the partitioning strategy, the data-partitioning and the space-partitioning based index structure. When the nearest neighbors search is applied on a data-partitioning index, additional clusters are visited due to the overlapping between the bounding forms (spheres or rectangles). In the case of the space-partitioning index; consultation of few populated or empty clusters is extremely probable. By using NOHIS-tree, the overlapping is avoided and the quality of clusters is preserved.

NOHIS indexing algorithm proceeds as follows:

1. The entire set of descriptors extracted from the image database constitutes the initial cluster; this cluster is divided into two sub-clusters using the hierarchical clustering algorithm PDDP [6]. Each of the two sub-clusters is divided into two partitions recursively. The result of the recursive division is a hierarchical structure of clusters arranged into a binary tree.
2. Descriptors of each obtained sub-clusters are gathered by hyper-rectangles directed according to the leading principal component to ensure the non-overlapping between the two bounding forms (see figure 2.d).

#### 2.1.2.1 Cluster partition

The entire set of descriptors is represented by an *n x m* matrix *M = ($d_1$, $d_2$, ..., $d_m$)* where each $d_i$ is a descriptor, *m* is the size of descriptors and *n* their dimension. Let *COV*, given by (1), be the covariance matrix and *U* its first principal component. Partition is made by the hyper-plane orthogonal to the leading principal component and passing through the centroid *w* of the cluster (see figure 2.b). The principal direction is the eigenvector corresponding to the largest eigenvalue of the covariance matrix.

$$COV = (M - we^T)(M - we^T)^T \quad (1)$$

$$w = \frac{1}{m}.(v_1 + v_2 + \cdots + v_m) = \frac{1}{m}.M.e \quad (2)$$

$$e = (1,1,\ldots,1)^T$$

Data is divided recursively into two parts $P_R$ and $P_L$ (R for right and L for left) according to the following rule:

$$g(d_i) = U^T(d_i - w) \geq 0 \quad \Rightarrow \quad d_i \in P_R$$

$$g(d_i) = U^T(d_i - w) < 0 \quad \Rightarrow \quad d_i \in P_L$$

Figure 2 illustrates the example of a cluster partition, in 2D, into two sub-clusters.

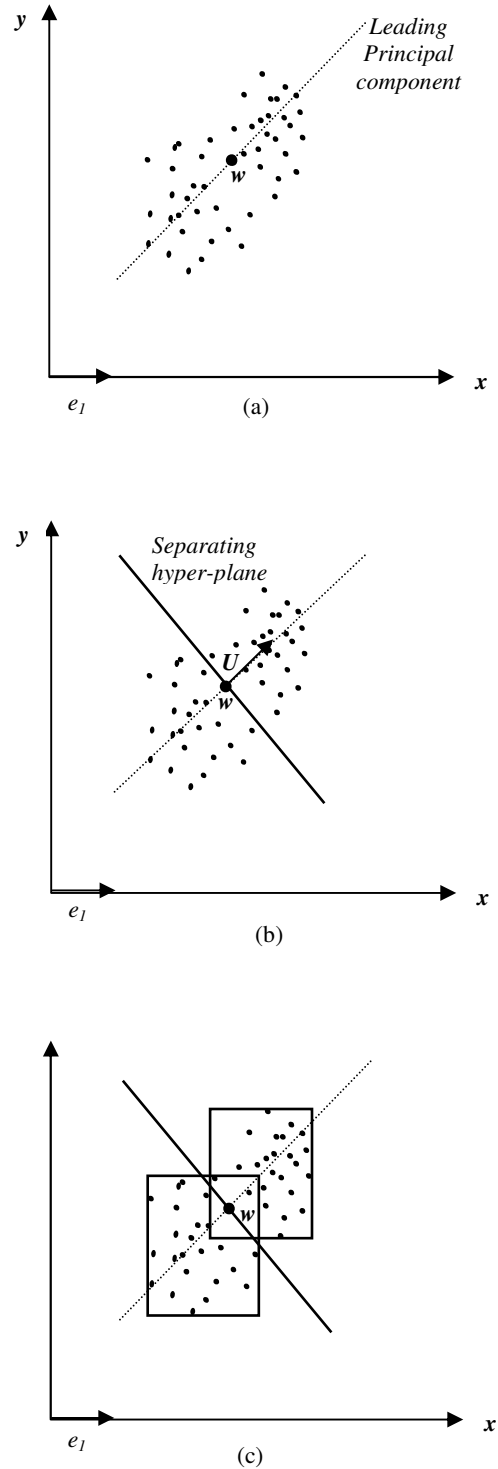

(a)

(b)

(c)

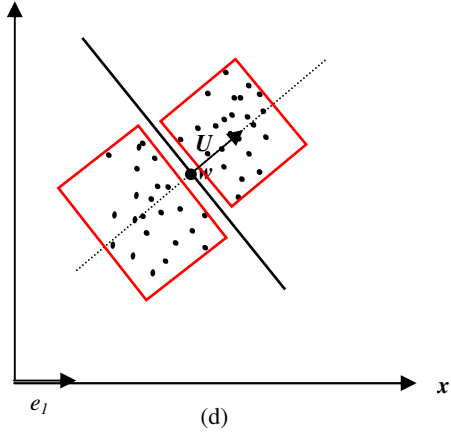

**Figure 2. Example, in 2D, of the indexing technique**

The algorithm of the clustering algorithm [7] is given in figure 3 below.

```
0.  Start with the matrix of vectors M(n x m), and
    a desired number of clusters c_max.
1.  Initialize Binary tree with a single Root node
2.  For c = 2,3,…., c_max do
3.      Select node C with largest ScatterValue
4.      Create L & R := left & right children of C
5.      For i = 1 to C_size
            Compute g(x_i), if g(x_i) ≤ 0 assign x_i to L
            else assign it to R
6.  Result: A binary tree with c_max leaf nodes
```

**Figure 3. PDDP Algorithm**

*2.1.2.2 Orientation of the bounding forms*

The bounding forms used to envelop the descriptors of clusters are the hyper-rectangles (starting from 3D). For clarity we will use the term MBR, for minimum bounding rectangle, to refer to the hyper-rectangle. The figure 2.c represents the case when the MBRs are in the origin reference mark; in which the coordinates of the vectors are expressed. It is clear that there is an overlap between the MBRs and consequently, in a nearest neighbors search, additional clusters will be visited without improving the results. To avoid the overlap, NOHIS indexing algorithm directs the MBRs according to the leading principal component (figure 2.d). In this case, a change of reference mark is essential to compute the coordinates of the descriptors in the new reference mark.

Let $B=\{e_1,e_2,…,e_n\}$ be the canonical base of $R^n$, $e_1=(1,0,0...,0)$, $e_2=(0,1,0...,0)$, $e_3=(0,0,1...,0)$, ….

The goal is to build an orthonormal base $B'=\{u_1,u_2,…,u_n\}$ where a vector is equal to $U$ $(u_1 = U)$, such a base can be obtained by transforming $B$ by an orthogonal isomorphism, for example by an orthogonal symmetry $S$. We must have $B'=(S(e_1),S(e_2),….,S(e_n))$ and in particular $S(e_1)= u_1=U$.

$$\text{Let } \quad V = \frac{(U-e_1)}{\|U-e_1\|} \quad (3)$$

and $H$ be the hyper-plane orthogonal to $V$, $H = V^\perp$, so that $H$ is the mediator hyper-plane of $e_1$ and $U$.

We define $S$ as the orthogonal symmetry with respect to $H$.

The image of the vector $x$ by $S$ is: $S(x) = x – 2<x,V>.V$, where $<x,V>$ is the scalar product of $x$ and $V$.

In particular, we have: $u_i = e_i – 2<e_i,V>.V$ $\quad 1 \le i \le n$

With this definition of $S$ when has, in fact, $u_1 = S(e_1) = U$.

In fact: $u_1 = e_1 - 2\alpha < e_1,U - e_1>.(U - e_1)$

With: $\quad \alpha = \frac{1}{\|U- e_1\|^2} = \frac{1}{\|U\|^2+\|e_1\|^2-2\langle e_1,\ U\rangle}$

$$\|U\|^2 = \|e\|^2 = 1 \Rightarrow \alpha = \frac{1}{2(1 - \langle e_1,\ U\rangle)}$$

$$u_1 = e_1 - \frac{2}{2(1 - \langle e_1,\ U\rangle)}(\langle e_1,\ U\rangle - 1)(U - e_1)$$

$$u_1 = e_1 + \frac{1}{(\langle e_1,\ U\rangle - 1)}(\langle e_1,\ U\rangle - 1)(U - e_1)$$

$$u_1 = e_1 + (U - e_1) = U$$

Let $N_R$ (resp. $N_L$) be the matrix containing the descriptors of $P_R$ (resp. $P_L$) in the base $B'$.

We have: $N_R = M_R - 2.V^T.V.\ M_R = (I - 2V^T.V).\ M_R$ $\quad (4)$

$N_L= M_L - 2.V^T.V.\ M_L = (I - 2V^T.V).\ M_L$ $\quad (5)$

Descriptors of $N_R$ (resp. $N_L$) are included in a MBR $R_R$ (resp. $R_L$). A property of the MBR is that each of his face passes by a descriptor at least. MBRs are characterized by the vectors $S$ and $T$, where:

$S_R = min\ (N_R)$ (resp. $S_L = min\ (N_L)$) $\quad$ and

$T_R = max\ (N_R)$ (resp. $T_L = max\ (N_L)$)

Note that the minimum and the maximum of this formula are taken line by line, so that $S = (s_1,..,s_i,...s_n)$ et $T = (t_1,..,t_i,...t_n)$ where $s_i$ (resp. $t_i$) is the minimum (resp. the maximum) of the $i^{th}$ component of the considered vectors.

The final result is a not-balanced binary tree called NOHIS-tree. In an internal node (not a leaf) of NOHIS-tree, following information are stored: $S_R$, $T_R$, $S_L$, $T_L$ and the common vector $V$ given by (3). A leaf node contains the descriptors. Leaves represent the obtained clusters.

## 2.2 On-line phase

The image retrieval process consists of querying the image database. When a user issues a query image to the system, this involves the following steps:

1. Computation of the descriptors of the query. Descriptors are computed by the same method used in the off-line phase.

2. Search the k-nearest neighbors for each descriptor in NOHIS-tree. This step returns the nearest neighbors and the ID of the images they belong to, which provides a list of images that may be similar to the query. The search algorithm will be described.

3. Matching the images obtained from the previous step with the query and display the images that the system considers similar to the query. The matching process is not explained in this paper.

### 2.2.1 The search algorithm

Before starting the search of nearest neighbors of a descriptor $q$, its coordinates in the new reference mark (i.e. the new base $B'$) must be computed in order to compute its distance to the MBR. The computing of new coordinates is done in each level in the NOHIS-tree until a leaf node. The passage of the $q$ from a father node to its child requires the computing of its new coordinates because a change of the reference mark has occurred. Two children of the same father have a common reference mark.

$B'$ is orthonormal, so coordinates of $q$ in $B'$ ($q'$) are given by the products scalar:

$$\langle q, u_i \rangle = \langle q, e_i \rangle - 2 \langle e_i, V \rangle \cdot \langle q, V \rangle$$

$$q' = [\langle q, u_1 \rangle, \langle q, u_2 \rangle, \ldots, \langle q, u_n \rangle]^T$$

$$q' = q - 2\langle q, V \rangle \cdot V \quad (6)$$

Distance separating $q$ from a rectangle $R$ is calculated as given in [8] using $q'$, $S$ and $T$. *MINDIST* is the distance between the query vector and an MBR.

$$MINDIST(q', R) = \sum_{i=1}^{n} |q'_i - r_i|^2 \quad (7)$$

$$\text{with:} \quad r_i = \begin{cases} s_i & \text{if } q'_i < s_i \\ t_i & \text{if } q'_i > t_i \\ q'_i & \text{else} \end{cases}$$

And when $q'$ is inside the rectangle $R$, $MINDIST(q', R) = 0$

In algorithm 1, we present our k-nearest neighbors search (k-nn) adapted to the obtained NOHIS-tree. We note that NOHIS-tree support also a range query search. For a vector query $q$, the k nearest vectors must be returned. *list_Neighbors* (LN) is the table containing k-nearest neighbors. For each nearest vector, LN must contain: its index in the database, the index of the cluster to which it belongs, and its distance from $q$. Distances of the returned nearest neighbors are initialized in the infinite value. Returned LN is sorted according to the distances. This algorithm is recursive; the first call is done with the root of the NOHIS-tree and a distance called maxDist initialized at 0. If the node is not a leaf then, first $q'$ is calculated by (6) and then distances MINDIST given by (7) are computed between $q'$ and the two children's MBRs of the node. A first recursive call in the algorithm 1 can be made with the child node having smallest distance MINDIST, let M[j] be this smallest distance.

We attribute to M[j] the maximum between M[j] itself and maxDist. The condition of this recursive call is that M[j] must be lower than the biggest distance contained in LN (LN.dist[k]). A second call can be made with the second child if the same condition is satisfied. Else if the considered node is a leaf, Euclidian distances between $q'$ and all the vectors of the node are computed. Only vectors having a distance lower than LN.dist[k] are inserted in LN by an insertion sort algorithm.

---

**Algorithm 1 :** *K-NN Search*

---

1. **Begin**
2.   **If** the node is a leaf
3.     **For** $i := 1$ to *node.size*
4.       Compute distance between *vectQuestion* and *node.vect*[i], let be *dist*;
5.       **if** (*dist* < *list_Neighbors.dist[k]*)
6.         Insertion sort of current vector in *list_Neighbors*
7.       **end if**
8.     **end For**
9.   **Else**
10.     **For** $j := 1$ to 2 (the two child nodes)
11.       Compute coordinates of *vectQuestion* in the new reference mark, let be *vectQuestion'*
12.       M[j] = *MINDIST*(*vectQuestion'*, MBR of *node.child j*)
13.     **end For**
14.     Take the child node having the smallest distance M[j];
15.     M[j] = max (maxDist, M[j])
16.     **if** (M[j] < *list_Neighbors.dist[k]*)
17.       Recursive call of *K-NN Search* passing the child node and M[j] as maxDist
18.     **end if**
19.     let MS the MINDIST of the second child
20.     MS = max(maxDist, MS)
21.     **if** (MS < *list_Neighbors.dist[k]*)
22.       Recursive call of *K-NN Search* with the second child and MS as maxDist
23.     **end if**
24.   **end Else**
24. **End**

---

The condition of the recursive call in algorithm 1 (M[j] < LN.dist[k]) is necessary because distances of vectors included in a MBR from a query vector can be only higher or equal to M[j], and as LN is sorted in the ascending order, therefore LN.dist[k] is the biggest distance contained in LN, and consequently if M[j] is not lower than LN.dist[k], the MBR cannot contains closer vectors to the query vector that those already found.

In a hierarchical index the bounding forms of a level are contained in that of the inferior level. Taking the example of a *father* node, the bounding forms of its children are contained in its bounding form and consequently, the distance from a query vector to the *father* node is lower or equal to its distances to the children. This gives a property to the hierarchical index that the distance of a query vector $q$ to the bounding forms increases from a level to that highest.

In our index structure NOHIS-tree, bounding rectangles of children ($R_1$, $R_2$) are not included completely in the bounding rectangle of their *father* node R, as shown in figure 4.

The instruction M[j] = max (maxDist, M[j]) in the line 15 of algorithm 1, (resp. MS = max (maxDist, MS) in the line 20), preserve the property that the distance increases from a level to that highest in the search tree. M[j] expresses the distance to their intersection.

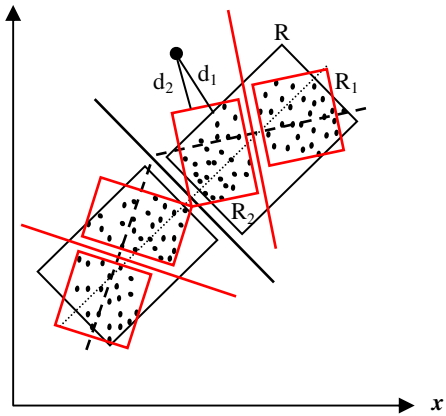

**Figure 4. Example of a rectangle with its children**

## 3. EXPERIMENTS

Descriptor computation algorithm, clustering algorithm and the search algorithm were implemented in C++. Algorithms run on a PC with Intel processor, its CPU is 1.8 GHz and 2 Go of RAM. We used *ImagEval* database for CBIR system evaluation, it consists of 9811 images. 2,416,975 descriptors were computed from these images.

Search time is an important factor to evaluate the performance of the CBIR system. In experiment 1 we compare three CBIR systems; NOHIS-Search system, the second is that using PDDP indexing algorithm and the third system is that using the sequential search. Table 1 and Figure 5 show the total search time for the three systems. The given times are the mean times when searching similar images for 10 queries; they include the time needed for the descriptor computation of the 10 queries. NOHIS-Search system significantly outperforms the other two systems. It performs the queries 19 times faster than the system using sequential search when using the database of 3800 images and 36 times faster than the system using sequential search when using the database of 9811 images. Besides, NOHIS-Search system is 2 times faster than the system using PDDP algorithm.

**Table 1. Mean time for the three systems**

| Size of the database | Mean time of search | | |
|---|---|---|---|
| | NOHIS-Search system | PDDP system | Seq.system |
| 3800 | 26,05 | 45,97 | 500 |
| 9811 | 50,15 | 106,98 | 1380 |

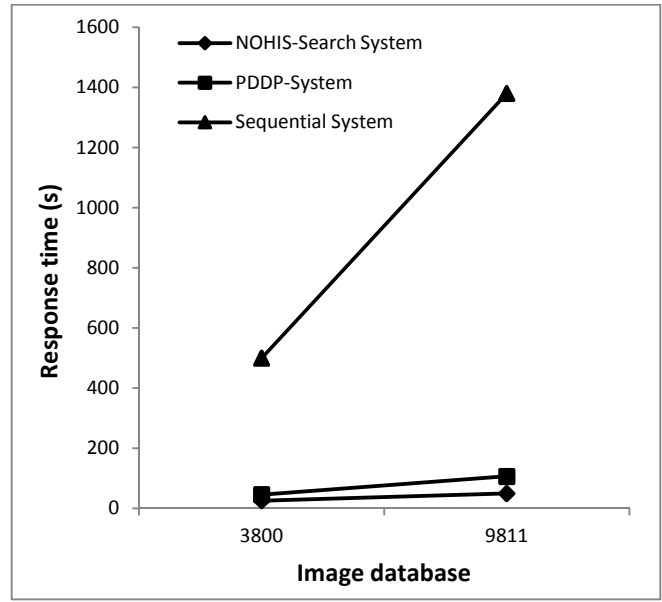

**Figure 5. Exp. 1, Retrieval time, 10 query images**

In experiment 2, the rapidity of NOHIS-Search system is explained. The number of the visited clusters is computed when comparing NOHIS-tree and PDDP-tree. Results given in table 2 and figure 6 show that less clusters are visited using NOHIS-tree than PDDP-tree. The orientation of MBRs in NOHIS-tree avoids the overlap which explains the obtained results.

**Table 2. The visited clusters when comparing NOHIS-tree and PDDP-tree**

| Size of the database | Number of clusters | Number of the visited clusters | |
|---|---|---|---|
| | | NOHIS-tree | PDDP-tree |
| 50,000 | 499 | 20 | 231 |
| 100,000 | 1189 | 24 | 529 |
| 250,000 | 1993 | 37 | 937 |
| 350,000 | 2397 | 41 | 1081 |
| 500,000 | 2995 | 43 | 882 |

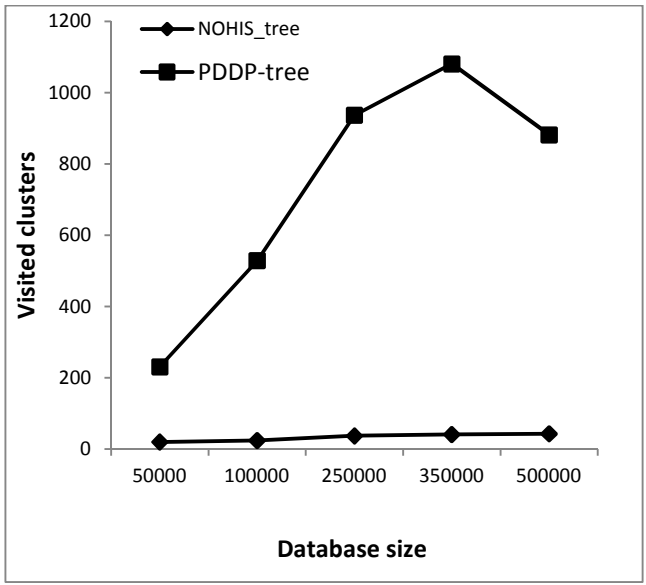

**Figure 6. Exp. 2, Visited clusters, 20 NNs for 200 query descriptors, increasing size**

Examples of images retrieved when using NOHIS-Search system are shown in the figures 7. In each figure, the first image in the top is the query and the other images are the responses of the system. Just the first 10 retrieved images are displayed.

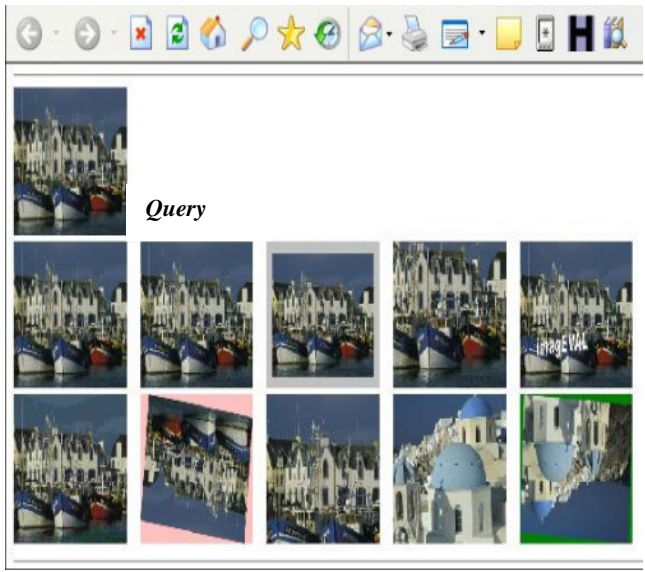

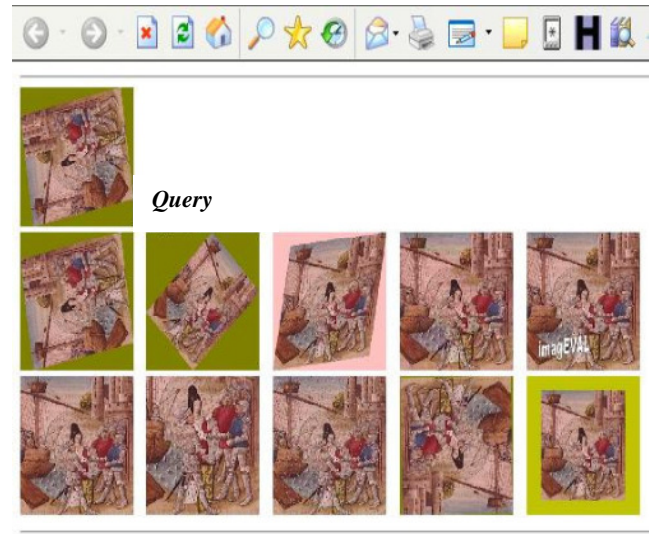

**Figure 7. Results with NOHIS-Search system**

## 4. CONCLUSION

A content-based image retrieval system called NOHIS-Search was presented in this paper, the system is based on the indexing technique NOHIS-tree. The on-line and off-line phases of the system were described. The performance evaluation of the proposed system with other systems shows that NOHIS-Search is faster that the two other systems. NOHIS-Search, however, requires further investigations especially in the matching process.

## 5. REFERENCES

[1] Faloutsos, C. 1996. Searching Multimedia Databases by Content. *Kluwer Academic Publishers*.

[2] Taileb, M. Lamrous, S., and Touati, S. 2008. Non Overlapping Hierarchical Index Struture. *In International Journal of Computer Science*, vol. 3 no. 1, pp. 29-35.

[3] Mikolajczyk, K., Schmid, C. 2004. Scale and affine invariant interest point detectors. *In International Journal of Computer Vision*. 60(1), pages : 63–86.

[4] Harris, C.J., Stephens, M. 1988. A combined corner and edge detector. *In Proceedings of the 4th Alvey Vision Conference*, Manchester, pages 147–151.

[5] Sayah, S.H.-K. 2007. Indexation d'images par moments: Acces par le contenu aux documents visuels. Doctoral Thesis. Ecole Normale Superieure de Cachan.

[6] D. L. Boley, D. L. 1998. Principal Direction Divisive Partitioning**.** *Data Mining and Knowledge Discovery* 2(4):325-344.

[7] Savaresi,S., Boley, D. L., Bittanti, S., Gazzaniga, G. 2000. *Choosing the cluster to split in bisecting divisive clustering algorithms*. CSE Report TR-00-055, University of Minnesota, 2000.

[8] Roussopoulos, N., Kelly, S., Vincent, F. 1995. Nearest Neighbor Queries". *In Proceeding of ACM SIGMOD,* May.